\documentclass[aps,prb,floatfix,superscriptaddress,twocolumn,footinbib]{revtex4-2}
\usepackage[colorlinks=true, linkcolor=blue, filecolor=blue, urlcolor=blue, citecolor=blue]{hyperref}
\usepackage{amsfonts}
\usepackage{bbm}
\usepackage{graphicx}
\usepackage{dcolumn}
\usepackage{bm}
\usepackage{soul}
\usepackage[version=3]{mhchem}
\usepackage{relsize}
\usepackage{mathrsfs}
\usepackage{ulem}  

\begin{document}

\title{Spin Splitting Nernst Effect in Altermagnet}

\author{Xing-Jian Yi}
\email[]{These authors contributed equally to this work.}
\affiliation{International Center for Quantum Materials, School of Physics, Peking University, Beijing 100871, China}
\affiliation{Hefei National Laboratory, Hefei 230088, China}

\author{Yue Mao}
\email[]{These authors contributed equally to this work.}
\affiliation{International Center for Quantum Materials, School of Physics, Peking University, Beijing 100871, China}

\author{Xiancong Lu}
\affiliation{Department of Physics, Xiamen University, Xiamen 361005, People's Republic of China}

\author{Qing-Feng Sun}
\email[]{sunqf@pku.edu.cn}
\affiliation{International Center for Quantum Materials, School of Physics, Peking University, Beijing 100871, China}
\affiliation{Hefei National Laboratory, Hefei 230088, China}
\affiliation{Collaborative Innovation Center of Quantum Matter, Beijing 100871, China}

\begin{abstract}
Altermagnet is a distinctive magnet phase, which has spin-split energy band but with zero net magnetic moment.
In this paper, we propose that altermagnet behaves spin splitting Nernst effect: Under a longitudinal temperature gradient, the electrons with opposite spins tend to split oppositely in the transverse direction, thus generating a transverse spin current.
The spin splitting Nernst effect is understood from the contribution of the longitudinal wave vector to the transverse group velocity.
Using the nonequilibrium Green's function method, we calculate the spin-dependent transmission coefficient in the four-terminal altermagnet device.
From the spin-dependent transmission coefficient, the nonzero transverse spin current from longitudinal temperature gradient is obtained, and the spin splitting Nernst effect is verified.
We systematically study the parameter dependence of the spin splitting Nernst effect, while also performing symmetry analysis.
The spin splitting Nernst effect can be easily regulated by Fermi surface energy, temperature, transport direction, and system size.
Furthermore, in altermagnet, the $xy$-response and $yx$-response spin splitting Nernst coefficients are equal with $N_{s,xy}=N_{s,yx}$, different from the conventional spin Nernst effect where they are opposite. Meanwhile, the spin splitting Nernst effect require neither spin-orbit coupling nor net magnetism.
\end{abstract}
\maketitle

\section{\label{sec1}Introduction}

Altermagnet is a novel magnet phase in condensed matter physics, which is characterized by a compensated antiparallel magnetic order in direct space with opposite-spin sublattices connected by crystal-rotation symmetries \cite{smejkal_Emerging_2022}.
Simultaneously, it breaks the time-reversal symmetry but keeps the space-inversion symmetry, leading to an anisotropic spin splitting with respect to the wave vector in the reciprocal momentum space 
\cite{smejkal_Emerging_2022, smejkal_Conventional_2022, hayami_MomentumDependent_2019}.
As a distinct and comparably abundant phase, altermagnetism emerges on the basic level of a nonrelativistic description of collinear magnets \cite{smejkal_Emerging_2022, smejkal_Conventional_2022}.
Altermagnet realizes $d$-wave magnetism, which is the magnetic counterpart of the unconventional $d$-wave superconductivity \cite{schofield_There_2009}.
Besides, altermagnet manifests interesting macroscopic response, such as the anomalous Hall effect \cite{han_Electrical_2024, feng_Anomalous_2022, smejkal_Crystal_2020}, charge-spin conversion\cite{bose_Tilted_2022}, giant and tunneling magnetoresistance effect \cite{smejkal_Giant_2022} and spin splitting \cite{bai_Observation_2022, gonzalez-hernandez_Efficient_2021, karube_Observation_2022}.
Based on the above advantages and its huge potential in spintronics, superconductivity and spin caloritronics, altermagnet has attracted wide attention
\cite{bai_Observation_2022, bai_Efficient_2023, bose_Tilted_2022, cheng_Orientationdependent_2024, gonzalezbetancourt_Spontaneous_2023, gonzalez-hernandez_Efficient_2021, karube_Observation_2022, ouassou_Dc_2023, papaj_Andreev_2023, samanta_Crystal_2020, shao_Spinneutral_2021, sun_Andreev_2023, zhou_Crystal_2024, cheng_Fieldfree_2024, amundsen_RKKY_2024, das_Crossed_2024, li_Realizing_2024, zeng_Description_2024}.
In addition, $\mathrm{RuO_2}$ \cite{bai_Observation_2022, feng_Anomalous_2022, smejkal_Crystal_2020}, $\mathrm{FeF_2}$ \cite{lopez-moreno_Firstprinciples_2012}, $\mathrm{MnF_2}$ \cite{yuan_Giant_2020} and $\mathrm{MnTe}$
\cite{lee_Broken_2024, mazin_Altermagnetism_2023}
have been proposed as the candidate materials for altermagnetism.

Spintronics, manipulating the spin degree of freedom to achieve the data storage and logical operations, has always been an active topic in physics in the past two decades \cite{sinova_New_2012, zutic_Spintronics_2004}.
Compared to charge, the regulation of spin degree has lower energy loss and faster reaction speed \cite{mao_Universal_2024, zutic_Spintronics_2004, fert_Nobel_2008, linder_Superconducting_2015}.
In spintronics, the generation of spin current lies at its heart.
One of the methods to generate spin current is the 
spin Nernst effect,
where a transverse spin current is induced by a longitudinal thermal gradient
\cite{cheng_Spin_2008, xing_Nernst_2009, kim_Observation_2017, wimmer_Firstprinciples_2013, bose_Recent_2019,  yang_Gate_2018,meyer_Observation_2017, sheng_Spin_2017,cheng_Spin_2016, kondo_Nonlinear_2022}.
The 
spin Nernst effect also offers the possibility to make devices combining heat and spin.
Recently, the 
spin Nernst effect has been widely explored in various systems by both theoretical and experimental studies.
For example, spin Nernst effect is reported by measuring differential voltage between two ferromagnet contacts placed on top of the $\mathrm{Pt}$ \cite{bose_Direct_2018}.
In addition, spin Nernst effect-induced transverse magnetoresistance in ferromagnet/non-magnetic heavy metal bilayers is experimentally observed \cite{kim_Observation_2017, meyer_Observation_2017, sheng_Spin_2017}.
On the theoretical side, various schemes for the spin Nernst effect have been proposed. The effect of extrinsic skew scattering and side jumps mechanism at impurities to spin Nernst effect is studied \cite{tauber_Extrinsic_2012, wimmer_Firstprinciples_2013}.
Impurities-independent intrinsic spin Nernst effect in various systems such as heavy metal \cite{salemi_Firstprinciples_2022}, quantum wells \cite{rothe_Spindependent_2012}, Dirac semimetals \cite{yen_Tunable_2020}, two-dimensional electron gas with a Rashba-type spin-orbit coupling \cite{dyrdal_Spin_2016}, antiferromagnets with the Dzyaloshinskii-Moriya interaction \cite{cheng_Spin_2016, zhang_SpinNernst_2018} and topological superconductors \cite{matsushita_SpinNernst_2022} are also theoretically predicted.
However, the 
spin Nernst effect is usually limited to demand the spin-orbit coupling \cite{manchon_Currentinduced_2019, sinova_Spin_2015,dyrdal_Spin_2016} or the participation of magnons \cite{zyuzin_Magnon_2016,cui_Efficient_2023}.
In the altermagnet, the electrons have a unique band structure with anisotropic momentum-dependent spin splitting.
Such spin-dependent anisotropy brings new possibilities in spin current generation,  especially spin splitter effect \cite{gonzalez-hernandez_Efficient_2021}.
Thus, altermagnet could be a platform to realize such an effect, without the requirement for spin-orbit coupling and magnons. 

\begin{figure}[!htb]
    \centerline{\includegraphics[width=\columnwidth]{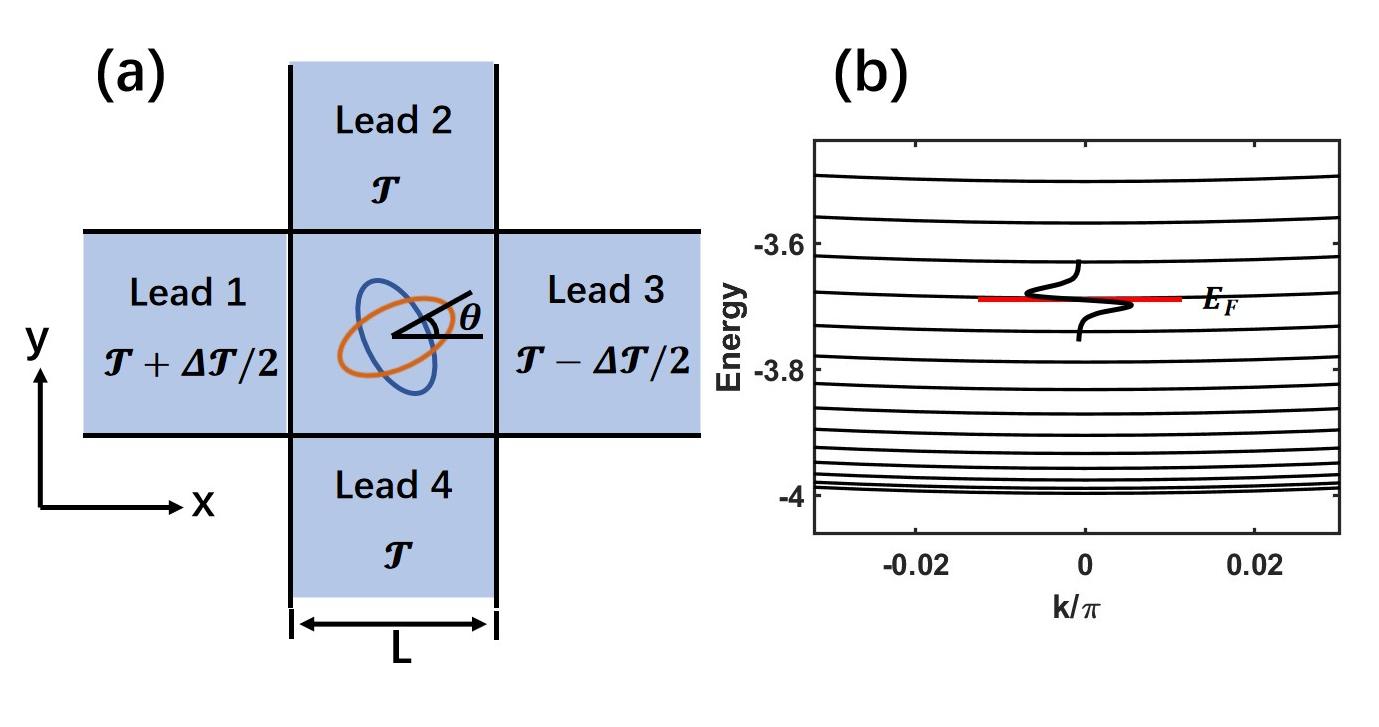}}
    \caption{(a) Schematic for four-terminal device with a longitudinal thermal gradient $\Delta \mathcal{T}$ between lead 1 and lead 3. The two ellipses in the central region denote Fermi surfaces for electrons with opposite spins.
    The orientation of altermagnet is drawn by the angle $\theta$
    between crystalline axis (the major axis of the red Fermi surface)
    and longitudinal direction. The size of the device is denoted by $L$.
    (b) Schematic view of the band structure of altermagnet nanoribbon.
    The black oscillatory line across $E_F$ is $f_0 (1-f_0)(E-E_F)$, which shows the contribution to spin splitting Nernst effect.}\label{Fig1}
\end{figure}

In this paper, we propose that the altermagnet behaves the spin splitting Nernst effect, where a longitudinal thermal gradient $\Delta\mathcal{T}$ induces a transverse spin current $J_{sH}$.
We call it spin splitting Nernst effect, because its mechanism is relevant to the spin splitter effect.
We also attribute the spin splitting Nernst effect to the transverse group velocities of electrons in altermagnet.
We verify the existence of spin splitting Nernst effect by studying a four-terminal altermagnet device, as schematically shown in Fig. \ref{Fig1}(a).
With the aid of the tight-binding model and the nonequilibrium Green's function method, the spin-dependent transmission coefficient and consequently the spin splitting Nernst coefficient $N_s$ $(N_s=J_{sH}/\Delta \mathcal{T})$ are obtained.
The nonzero spin splitting Nernst signal confirms the existence of the spin splitting Nernst effect in altermagnet.
Furthermore, we give a systematic study about the effects of parameters on the spin splitting Nernst effect.
At low temperature, the spin splitting Nernst coefficient versus Fermi energy behaves peaks due to the subbands as a quantum effect. 
With the increase of the temperature, the peaks are smoothed with a nonzero spin splitting Nernst coefficient left,
and the spin splitting Nernst coefficient is enhanced.
The spin splitting Nernst effect can be enhanced by increasing the temperature, the strength of altermagnet, or the system size.
The spin splitting Nernst coefficient has a quasi-sinusoidal relationship with the angle between crystalline axis and longitudinal direction.
Thus, the spin splitting Nernst effect can be effectively regulated by the Fermi energy, the temperature, the transport direction, and the system size.

The rest of the paper is organized as follows.
In Sec. \ref{sec2}, the mechanism of spin splitting Nernst effect is given,
the effective tight-binding Hamiltonian for four-terminal device is introduced,
and the formalisms for calculating the spin splitting Nernst coefficient are then derived.
Sec. \ref{sec3} gives the numerical results of transmission coefficient and the spin splitting Nernst coefficient along with discussions.
In Sec. \ref{sec4}, the effects of the temperature, strength of altermagnet, the angle between crystalline axis and longitudinal direction, and the size of device on the thermoelectric property are studied.
Finally, a discussion and summary are given in Sec. \ref{sec5}.

\section{\label{sec2}Mechanism and methods}

\begin{figure}[!htb]
\centerline{\includegraphics[width=\columnwidth]{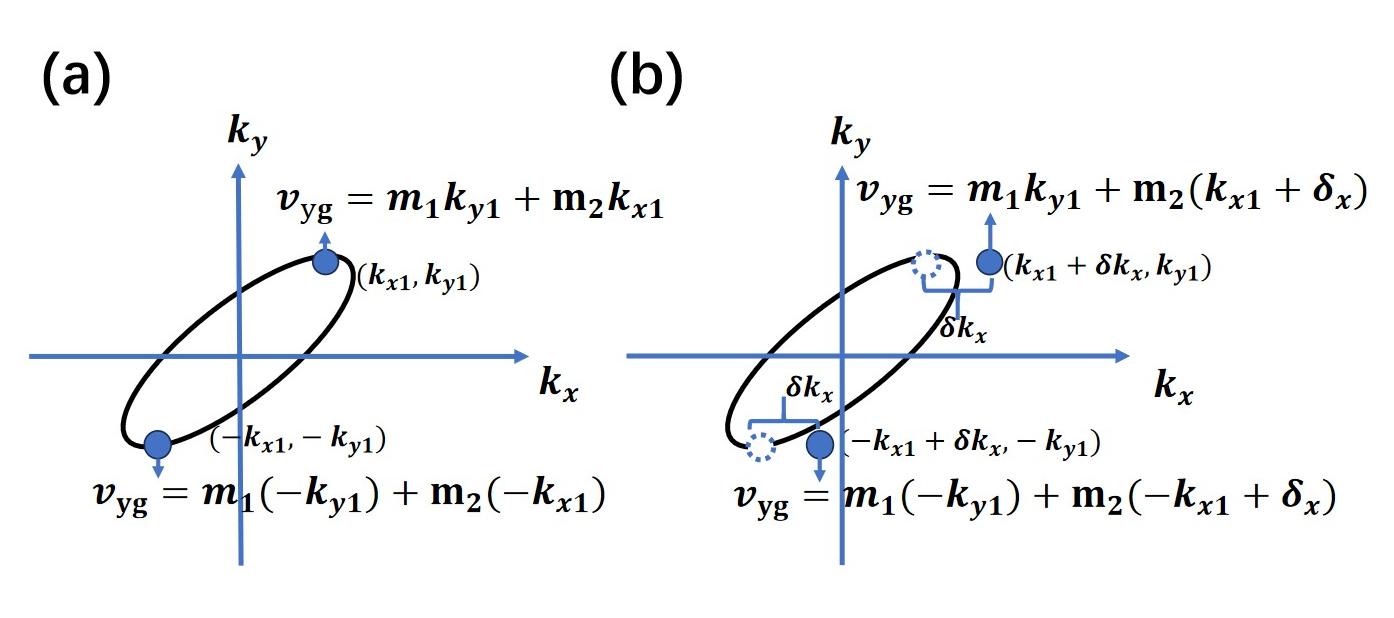}}
\caption{Physical picture for spin splitting Nernst effect. 
(a) Fermi surface of altermagnet at equilibrium.
The black ellipse is the cross-section of the energy band of spin-up electrons, with the two symmetric blue dots being the electrons with opposite wave vectors. 
The $v_{yg}$ is the transverse group velocity, which sums to be zero.
(b) Fermi surface of altermagnet under thermal drive.
The thermal gradient induces the drift wave vector $\delta k_x$, and a nonzero net transverse group velocity is generated.
}\label{Fig2}
\end{figure}

The Hamiltonian of the altermagnet in momentum space is written as
\cite{cheng_Orientationdependent_2024, cheng_Fieldfree_2024, wei_Gapless_2024}
\begin{align}
H&=\frac{\hbar^2 k^2}{2m^*}-V_0 \nonumber \\
&-\frac{\alpha_0}{2} [(k_x^2-k_y^2 ) \cos 2\theta+2k_x k_y \sin 2\theta ] \sigma_z, \label{E1}
\end{align}
where $k^2=k_x^2+k_y^2$.
Here $m^*$ is effective electron mass, $k_x,k_y$ are the wave vectors in $x$, $y$ directions, and $V_0$ is the potential energy.
Moreover, $\alpha_0$ is the parameter which measures the strength of the altermagnet.
$\theta$ is the angle between the crystalline axis and the longitudinal direction
[see Fig. \ref{Fig1}(a)].
$\sigma_z$ is the Pauli matrice in spin space.
We set $x$ axis as the longitudinal direction and $y$ axis as the transverse direction.

From this Hamiltonian, the mechanism for spin splitting Nernst effect can be derived.
First, considering the spin-up electron, its transverse group velocity is given by
\begin{align}
    v_y&=\frac{1}{\hbar}\frac{\partial H}{\partial k_y }=\frac{\hbar k_y}{m^*}
    -\frac{\alpha_0}{\hbar}(-k_y \cos2\theta+k_x \sin2\theta)    \nonumber \\
    &=m_1 k_y+m_2 k_x, \label{E2}
\end{align}
with
\begin{align}
    m_1=\frac{\hbar}{m}+\frac{\alpha_0}{\hbar}\cos2\theta, \ \ \  m_2=-\frac{\alpha_0}{\hbar} \sin2\theta     .\label{E3}
\end{align}
For a normal metal with $\alpha_0=0$, the Fermi surfaces are circular, and only $k_y$ contributes to the transverse group velocity $v_y$.
But for altermagnet with nonzero $\alpha_0$, the Fermi surfaces are elliptical.
From Eq. (\ref{E3}), the transverse group velocity is not only contributed by $k_y$ but also by $k_x$.
As shown in Fig. \ref{Fig2}(a), for a spin-up electron with the wave vector $(k_{x1},k_{y1})$ at the Fermi surface, we can always find its counterpart spin-up electron at $(-k_{x1},-k_{y1})$ which is symmetric with the former about $(0,0)$.
The two electrons each has a transverse group velocity, which is obtained by substituting $\pm (k_{x1},k_{y1})$ into Eq. (\ref{E2}) .
The two velocities compensate each other and sum to be zero.
As the whole system is composed of such electron pairs, the total transverse group velocity is also zero.

When a temperature gradient is applied in the longitudinal direction, the electrons
with the energy $E$ higher than Fermi energy $E_F$ tend to move from the high temperature region to the low temperature region.
This corresponds to a drift wave vector $\delta k_x$ in the longitudinal direction, as shown in Fig. \ref{Fig2}(b).
As such, the wave vectors of electrons are shifted to $(k_{x1}+\delta k_x,k_{y1})$ and $(-k_{x1}+\delta k_x,-k_{y1})$.
The sum of their transverse group velocities has a nonzero term $2 m_2 \delta k_x = -2 \frac{\alpha_0} {\hbar} \sin 2\theta \delta k_x$ [see Fig. \ref{Fig2}(b)].
For spin-down electrons, in the same way we obtain the net transverse group velocity, which is just opposite to that of spin-up electrons.
This indicates that the longitudinal thermal drive leads to opposite transverse motions for spin-up and spin-down electrons, i.e. the spin splitting Nernst effect appears.
Here, the effect can be called spin splitting spin Nernst effect
(spin splitting Nernst effect for short), as the thermo-driven analogy to the spin splitter effect \cite{gonzalez-hernandez_Efficient_2021}.
Compared with the conventional spin Nernst effect, the spin splitting Nernst effect originates from the anisotropic spin splitting in the altermagnet
and its coefficient has different symmetry (see the discussions in Sec. \ref{sec5}).
Under the combined contributions of electrons with opposite spins, a nonzero transverse spin current can be induced and is proportional to $\alpha_0 \sin2\theta$, which seems to be dependent on the altermagnet strength $\alpha_0$ and angle $\theta$.

On the other hand, under the temperature gradient,
the low-energy electrons with the energy $E$ lower than Fermi energy $E_F$ move
from the low temperature to the high temperature, which is opposite to high-energy electrons.
These low-energy electrons can induce a spin splitting Nernst effect opposite to that of high-energy electrons and partially cancel each other out.
However, the transmission coefficients are usually energy dependent,
and a net spin splitting Nernst effect still exists.

Next, we confirm this spin splitting Nernst effect by studying a four-terminal altermagnet device, as schematically shown in Fig. \ref{Fig1}(a).
It consists of a square center region connected to four semi-infinite leads, with a thermal gradient $\Delta\mathcal{T}$ applied longitudinally.
To discretize the Hamiltonian in Eq. (\ref{E1}),
the square-lattice Hamiltonian of the four-terminal altermagnet device can be obtained in the tight-binding representation
as\cite{cheng_Orientationdependent_2024, cheng_Fieldfree_2024, wei_Gapless_2024}
\begin{align}
    H_A&=\sum_{nm} \varepsilon_{nm}a_{nm}^\dagger a_{nm} + \sum_{nm} (a_{nm}^\dagger H_x a_{n+1,m}  \nonumber \\
    &+ a_{nm}^\dagger H_y a_{n,m+1}+ a_{nm}^\dagger  H_{xy} a_{n+1,m+1} \nonumber \\
    &+ a_{nm}^\dagger H_{x\overline{y}} a_{n+1,m-1}+H.c.) , \label{E4}
\end{align}
with
\begin{align}
    H_x&=-t(\sigma_0-\alpha\sigma_z \cos 2\theta), \ \ \ H_y=-t(\sigma_0+\alpha\sigma_z \cos 2\theta), \nonumber \\
    H_{xy}&=t \frac{\alpha}{2}  \sin 2\theta \sigma_z, \qquad\qquad H_{x\overline{y}}=-t \frac{\alpha}{2}  \sin 2\theta \sigma_z
    ,\label{E5}
\end{align}
where $a_{nm}^\dagger$ $(a_{nm})$ is the creation (annihilation) operator of electrons in the site $(n,m)$ with $n$ being the $x$ coordinate and $m$ being the $y$ coordinate, and in the electron basis $a_{nm}=(a_{nm\uparrow},a_{nm\downarrow})^T$.
$\sigma_0$ is the $2\times 2$ unit matrix.
$t=\hbar^2/(2m^*a^2)$ is the hopping matrix element with the lattice constant $a$.
$\varepsilon_{nm}=4t-V_0$ is the on-site energy which is fixed as zero by setting $V_0=4t$.
Here $\alpha$ is a dimensionless number and $\alpha=\frac{m^*}{\hbar^2} \alpha_0$. 
Using $a$ as length unit, the size of the central region is $L \times L$.
The particle current from the $p$-th lead with spin $\sigma$
flowing into the center region is calculated from the Landauer-B$\ddot{u}$ttiker formula
\cite{cheng_Spin_2008, yang_Gate_2018, mao_Charge_2021, yang_Linear_2020}
\begin{align}
    J_{p\sigma}=\frac{1}{\hbar} \sum_{q\neq p} \int \frac{dE}{2\pi} T_{p\sigma,q}(E) [f_p(E)-f_q(E)], \label{E6}
\end{align}
where $T_{p\sigma,q}$ is the transmission coefficient from lead $q$ to lead $p$ with spin $\sigma$ and $E$ is the energy of the incident electron. By using the nonequilibrium Green's function method, the transmission coefficient can be calculated as $T_{p\sigma,q}(E)=\mathrm{Tr}[\bm{\Gamma}_{p\sigma} \mathbf{G}^r \bm{\Gamma}_q \mathbf{G}^a]$,
in which $\bm{\Gamma}_{p\sigma}(E)=i(\bm{\sum}\nolimits_{p\sigma}^r-\bm{\sum}\nolimits_{p\sigma}^{r\dagger}), \bm{\Gamma}_q=\bm{\Gamma}_ {q\uparrow}+\bm{\Gamma}_{q\downarrow}$ are the linewidth functions. The Green's function is $\mathbf{G}^r(E)=[\mathbf{G}^a(E)]^\dagger=\{E\mathbf{I}-\mathbf{H}_0-\sum_p {\bm{\sum}}_p^r \}^{-1}$.
$\mathbf{H}_0$ is the Hamiltonian of the central region, $\bm{\sum}\nolimits_p^r$ is the retarded self-energy function originating from the coupling of the lead $p$, which can be calculated by $\bm{\sum}\nolimits_p^r=\bm\sum\nolimits_{p\uparrow}^r \oplus \bm\sum\nolimits_{p\downarrow}^r$.
Because the Hamiltonian is block diagonal, the self energy is obtained separately from spin $\sigma=\uparrow,\downarrow$ blocks.
$\bm\sum\nolimits_{p\sigma}^r$ is obtained by $\bm\sum\nolimits_{p\sigma}^r=H_{cp,\sigma}g_{p,\sigma}^r(E)H_{pc,\sigma}$,
where $g_{p,\sigma}^r(E)$ is the surface Green's function of lead $p$ for spin $\sigma$ \cite{lee_Simple_1981,sancho_highly_1985}, and $H_{cp,\sigma}$ is the coupling matrix from central region to lead $p$ for spin-$\sigma$ electrons. $H_{pc,\sigma}=H_{cp,\sigma}^\dagger$.
$f_p(E)$ is the electronic Fermi distribution function of lead $p$ and $f_p(E,E_F^p,\mathcal{T}_p )=1/\{ exp[(E-E_F^p )/k_B \mathcal{T}_P]+1\}$, which is the function of temperature $\mathcal{T}_p$ and the Fermi energy $E_F^p$.
$E_F^p=E_F+eV_p$, with $e$ the electron charge, $V_p$ the bias voltage, and $E_F$ the Fermi energy.
After getting the particle current $J_{p\sigma}$, we obtain the spin current $J_{ps}=(\hbar/2)(J_{p\uparrow }-J_{p\downarrow })$.
Considering that a small temperature gradient $\Delta \mathcal{T}$ and zero bias are added longitudinally between the lead 1 and lead 3, temperatures in the leads are $\mathcal{T}_1=\mathcal{T}+\frac{\Delta \mathcal{T}}{2},\mathcal{T}_3=\mathcal{T}-\frac{\Delta \mathcal{T}}{2},\mathcal{T}_2=\mathcal{T}_4=\mathcal{T}$, and the biases $V_1=V_3=0$.
Considering the closed boundary condition $(V_2=V_4=0)$ and in the case of a small thermal gradient, the Fermi distribution function can be expanded linearly
\cite{yang_Gate_2018}
\begin{align}
    f_p (E,\mathcal{T}_p )&=f_0+\Delta{\mathcal{T}_P} \frac{\partial{f}}{\partial \mathcal{T}_p} \bigg|_{V_p=0,\mathcal{T}_p=\mathcal{T}}	\nonumber \\
    &=f_0+f_0 (1-f_0 ) \times \left [(E-E_F )  \frac{\Delta {\mathcal{T}_P}}{k_B \mathcal{T}^2 } \right ] , \label{E7}
\end{align}
where $f_0=1/\ \{exp[(E-E_F)/k_B \mathcal{T}]+1\ \}$ is the Fermi distribution function at zero bias, and $\Delta {\mathcal{T}_P}=\mathcal{T}_P-\mathcal{T}$.
Then, the spin splitting Nernst coefficient $N_s=J_{2s}/\Delta \mathcal{T}$ is given by
\cite{cheng_Spin_2008}
\begin{align}
    N_s=\frac{1}{4\pi} \int dEf_0 (1-f_0) \left [\frac{(E-E_f )}{k_B \mathcal{T}^2} (\Delta T_{23}-\Delta T_{21} )\right ]  ,     \label{E8}
\end{align}
where $\Delta T_{2p}=T_{2\uparrow, p}-T_{2\downarrow,p}$. From Eq. (\ref{E8}), $N_s$ comes from the different deflecting destination of the spin-up and spin-down electrons, which can be represented by $T_{2\uparrow, 1} \neq T_{2\downarrow,1}, T_{2\uparrow, 3} \neq T_{2\downarrow,3}$.

\begin{figure}[!htb]
    \centerline{\includegraphics[width=\columnwidth]{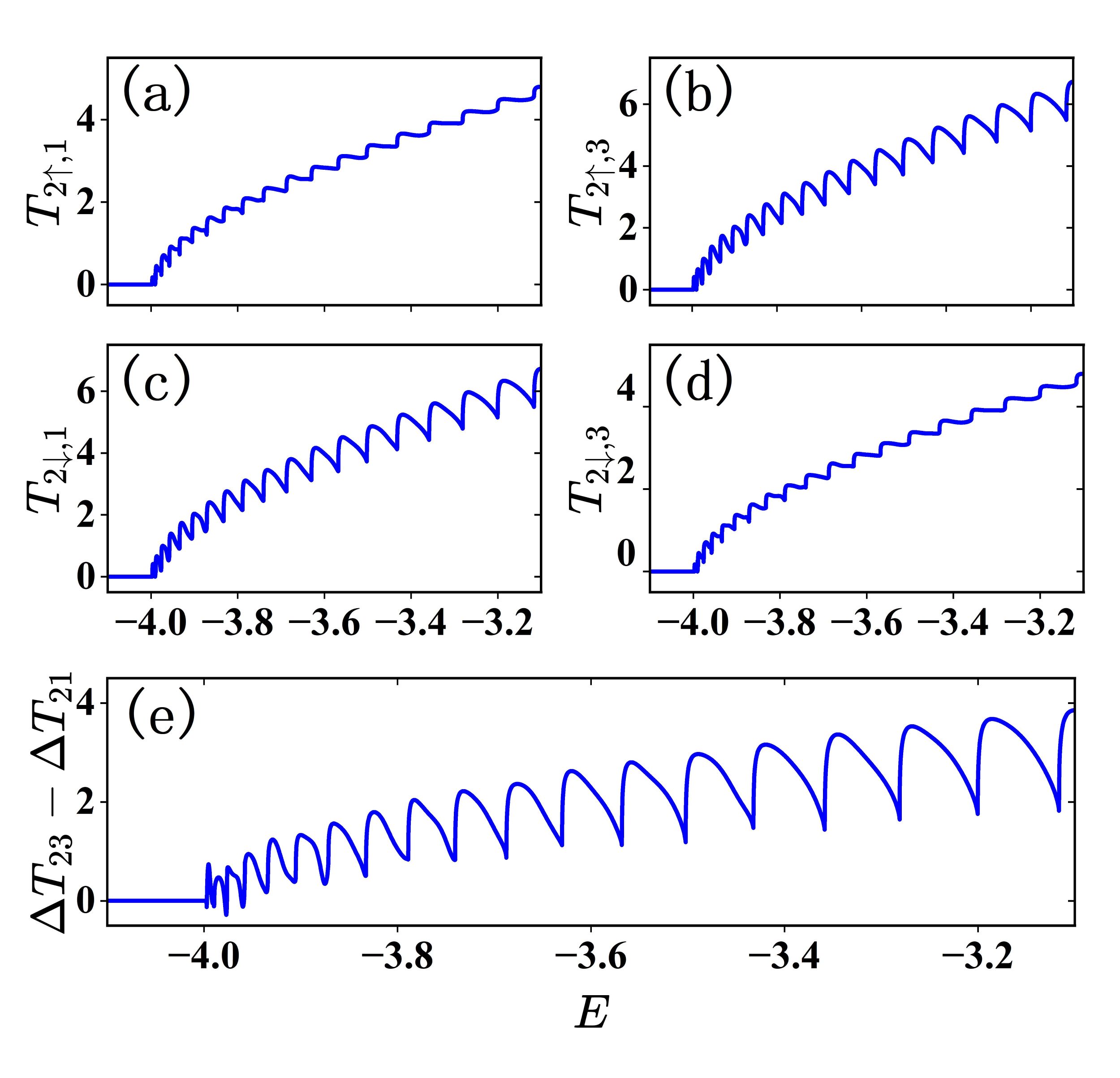}}
    \caption{(a-d) The four transmission coefficients versus the energy $E$
    of the incident electron. (e) $\Delta T_{23}-\Delta T_{21}$ versus $E$. The strength of altermagnet $\alpha=0.2$, the angle $\theta=\pi/4$, and the size of central region $L=60$.
    }\label{Fig3}
\end{figure}

In the numerical calculations, we set the hopping $t=1$, the lattice constant $a=1$, the Boltzmann constant $k_B=1$ as units.
If the effective electron mass is set as $m^*=0.05m_e$ ($m_e$ is the bare electron mass), the lattice constant is $a=1.35$ nm, $t$ will be about $0.42$ eV.
When the temperature is set at $\mathcal{T}=0.06$, it corresponds to about $290$ K, the room temperature.
When the temperature is set at $\mathcal{T}=0.001$, it corresponds to a low temperature about $5$ K.
In fact, in recent years, with the development of low temperature measurement technology, thermoelectric measurement of low dimensional samples at low temperature has also become feasible \cite{matsushita_Thermoelectric_2017, scheibner_Sequential_2007, shapiro_Thermoelectric_2017}.

\section{\label{sec3}Emergence of spin splitting Nernst effect}

In this section, we numerically show the appearance of the spin splitting Nernst effect
in altermagnet from the transmission coefficients.
In Figs. \ref{Fig3}(a-d), the four transmission coefficients related to the spin splitting Nernst coefficient are displayed.
They present a series of oscillatory structures, which just corresponds to the energy-band structure of altermagnet nanoribbon shown in Fig. \ref{Fig1}(b).
When the energy crosses the bottoms of the subbands, the number of the transport modes is increased \cite{datta_Electronic_1995}, thus the transmission coefficient abruptly jumps in Figs. \ref{Fig3}(a-d).
The Hamiltonians in Eq. (\ref{E5}) are spin-dependent, so the Green's functions and self energies for different spins are different, and the transmission coefficients are different for different spins $T_{2\uparrow,1} \neq T_{2\downarrow,1}$, $T_{2\uparrow,3} \neq T_{2\downarrow,3}$.
Also, the coefficients satisfy $T_{2\uparrow,1}\neq T_{2\uparrow,3}$, $T_{2\downarrow,1}\neq T_{2\downarrow,3}$.
This means both spin-up and spin-down electrons have different probabilities to move from lead 1 or lead 3 to lead 2.
However, $T_{2\uparrow,1}=T_{2\downarrow,3}$, $T_{2\downarrow,1}=T_{2\uparrow,3}$, which indicates that the motion tendency for spin-up and spin-down electrons is opposite.
Thus, $\Delta T_{23}-\Delta T_{21}$ in Fig. \ref{Fig3}(e) is nonzero and leads to a nonzero spin splitting Nernst coefficient, according to Eq. (\ref{E8}).

In fact, the relations $T_{2\uparrow,1}=T_{2\downarrow,3}$, $T_{2\downarrow,1}=T_{2\uparrow,3}$
can also be obtained by using the time-reversal operation $\mathscr{T}$ and the rotation operation about the $z$ axis $R_z (\pi/2)$ \cite{cheng_Spintriplet_2021, fang_Bulk_2012, yan_Anomalous_2020}.
The effects of the operations on the Hamiltonian are given by
\begin{align}
    &\mathscr{T}H(\alpha,\theta) \mathscr{T}^{-1}=H(-\alpha,\theta), \nonumber \\
    &R_z (\frac{\pi}{2})H(\alpha,\theta) R_z (\frac{\pi}{2})^{-1}=H(-\alpha,\theta) .        \label{E9}
\end{align}
Under the joint operation $\mathscr{T} R_z (\pi/2)$, the Hamiltonian keeps unchanged, but the transmission processes are changed:
For the process that spin-up electrons flow from lead 1 to lead 2, under this joint operation, it is changed to that spin-down electrons flow from lead 3 to lead 2.
Thus, $T_{2\uparrow,1}=T_{2\downarrow,3}$.
On the other hand, $T_{2\downarrow,1}=T_{2\uparrow,3}$ is obtained in the same way.
Note that here the nonzero altermagnet strength $\alpha$ is necessary for the spin splitting Nernst effect.
On the contrary, when $\alpha=0$ as a normal metal, the Hamiltonian satisfies $\sigma_x H(\alpha,\theta) \sigma_x ^{-1}=H(-\alpha,\theta)=H(\alpha,\theta)$ under operator $\sigma_x$ that exchanges spin up and down.
Thus, the Hamiltonian keeps unchanged, and any process of the spin-up electrons turns to be the same as the spin-down electrons, leading to the disappearance of
the spin splitting Nernst effect.

\begin{figure}[!htb]
\centerline{\includegraphics[width=\columnwidth]{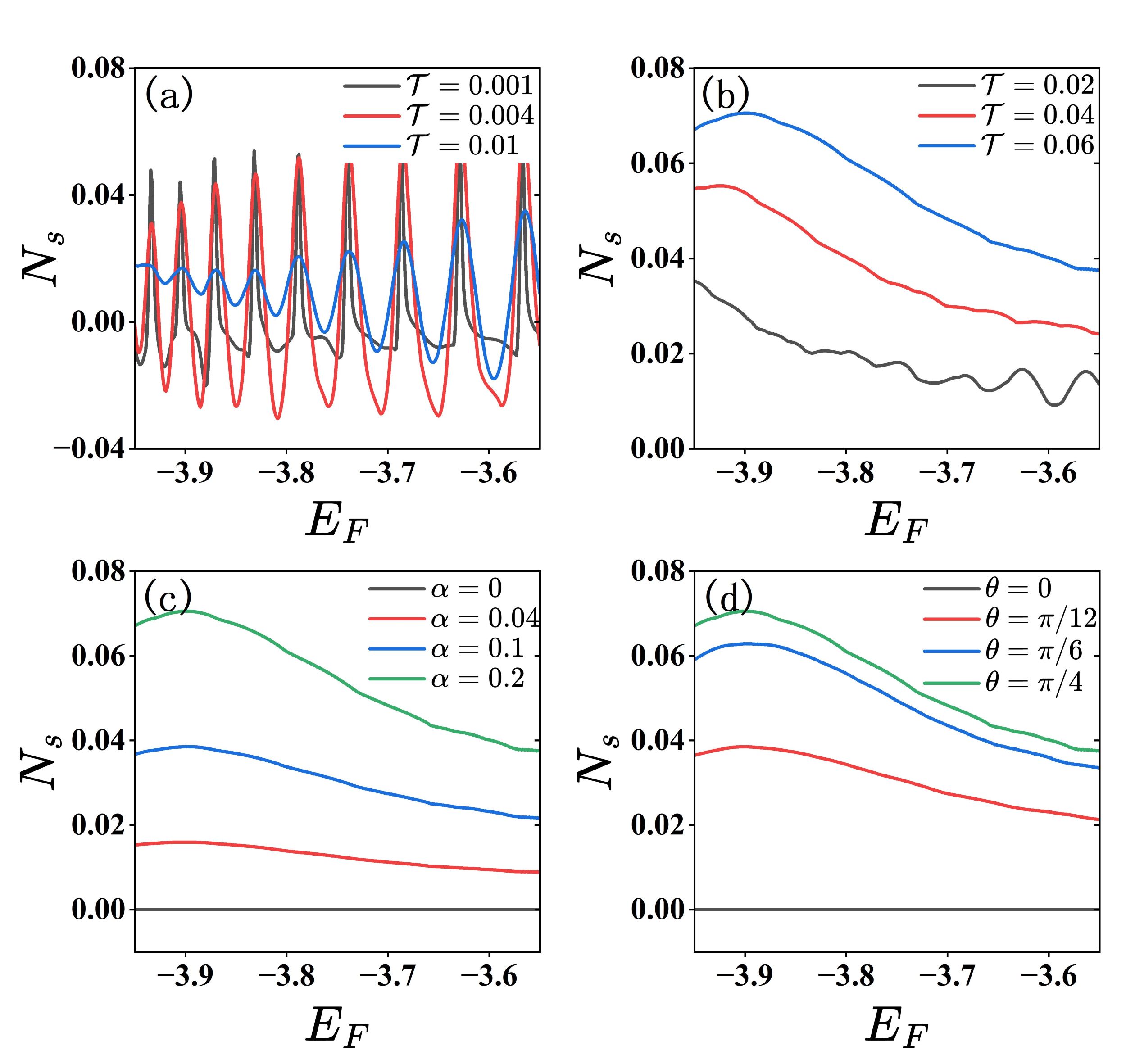}}
\caption{
The spin splitting Nernst coefficient $N_s$ versus Fermi energy $E_F$ for different parameters.
The parameters are: $\alpha=0.2$ in (a, b) and (d), $\mathcal{T}=0.06$ in (c, d), $\theta=\pi/4$  in panels (a-c), and $L=60$ in (a-d).}\label{Fig4}
\end{figure}

By performing the integral in Eq. (\ref{E8}), the spin splitting Nernst coefficient $N_s$ is obtained.
As shown in Fig. \ref{Fig4}, we indeed obtain the nonzero spin splitting Nernst coefficient, which confirms the existence of spin splitting Nernst effect in altermagnet.
Figs. \ref{Fig4}(a, b) show the spin splitting Nernst coefficient $N_s$ versus the Fermi energy $E_F$ at different temperatures.
At the low temperature in Fig. \ref{Fig4}(a), $N_s$ exhibits a series of peaks.
$N_s$ peaks when $E_F$ locates at the bottoms of the subbands because of the corresponding sudden jumps of the transmission coefficients in Fig. \ref{Fig3}, and it is damped when $E_F$ lies between bottoms of adjacent subbands. 
Let us explain the characteristic with the aid of the energy band in Fig. \ref{Fig1}(b).
From Eq. (\ref{E8}), $f_0 (1-f_0)(E-E_F)$ is the weight of the integral (also plotted in Fig. \ref{Fig1}(b)).
It exhibits an oscillatory structure around $E_F$, where the electrons with
incident energy $E$ above and below $E_F$ contribute opposite signs to the transverse spin current \cite{bose_Direct_2018}.
Therefore, the spin splitting Nernst coefficient $N_s$ is approximately proportional to the derivative $[\frac{d(\Delta T_{23}-\Delta T_{21})}{dE}] \big|_{E=E_F}$ at low temperature \cite{cheng_Spin_2008}.
When the incident energy $E$ crosses the bottom of a subband,
$\Delta T_{23}-\Delta T_{21}$ displays a sudden jump, as shown in Fig. \ref{Fig3}(e).
Thus, its derivative is maximized, and a peak of spin splitting Nernst coefficient $N_s$ is induced
when $E_F$ locates at the bottoms of the subbands.
On the other hand, the effective range of the integral is proportional to the temperature $\mathcal{T}$.
With the increase of the temperature, the oscillatory structure of $N_s$ is damped, as shown in Figs. \ref{Fig4}(a, b).
As the temperature further increases and exceeds the spacing between adjacent subbands, the $N_s$ becomes a smooth curve with nonzero value (such as $\mathcal{T}=0.04, 0.06$)  \cite{xing_Nernst_2009}.
Also, the increase of temperature involves more electrons into the integral, thus increasing the overall spin splitting Nernst coefficient.
In general, $N_s$ becomes lower as the $E_F$ is raised, except when $E_F$ is set around the bottom ($-4t$) of whole energy band. That is because of the lack of contributing electrons below the band bottom.
Fig. \ref{Fig4}(c) displays the $N_s$ versus $E_F$ at different strengths of altermagnet $\alpha$, and Fig. \ref{Fig4}(d) shows the $N_s$ versus $E_F$ at different angles $\theta$ between crystal axis and longitudinal direction.
From the analytic derivation in Sec. \ref{sec2}, the emergence of spin splitting Nernst effect demands a finite altermagnet strength $\alpha$ and an appropriate angle $\theta$.
Thus, when $\alpha$ or $\theta$ is zero, there is not the spin splitting Nernst effect. 
Within a certain range, increasing $\alpha$ or $\theta$ from zero can promote the spin splitting Nernst effect.

\begin{figure}[!htb]
\centerline{\includegraphics[width=\columnwidth]{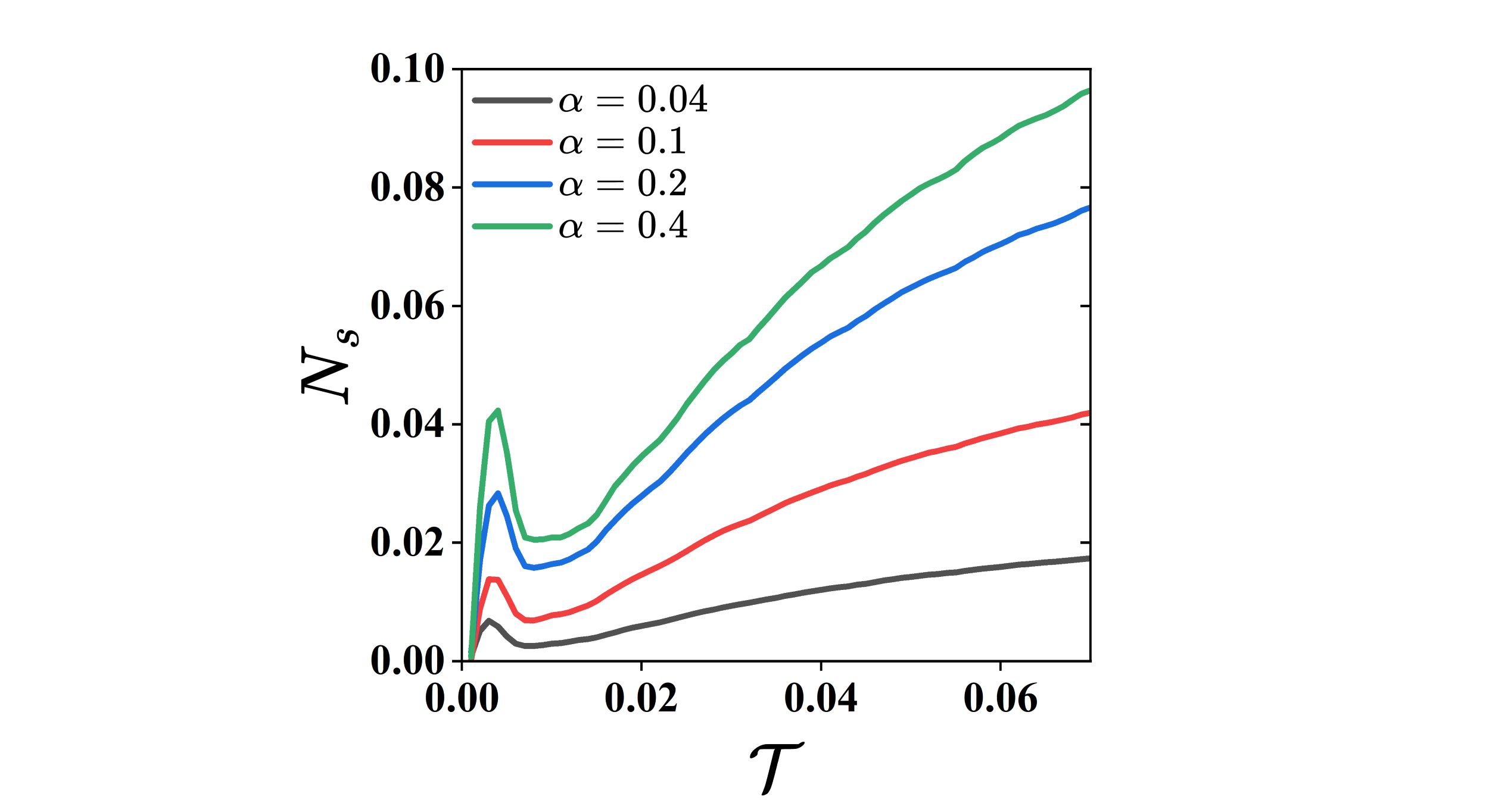}}
\caption{$N_s$ versus $\mathcal{T}$ for different $\alpha$. The parameters are the Fermi energy $E_F=-3.9$, $\theta=\pi/4$, and $L=60$. }\label{Fig5}
\end{figure}

\section{\label{sec4}Regulation on spin splitting Nernst effect}

In this section, we thoroughly study the effects of the temperature $\mathcal{T}$, the strength of altermagnet $\alpha$, the angle $\theta$, and the size of the sample $L$ on the spin splitting Nernst coefficient $N_s$.

First, we study the temperature effect. We plot the spin splitting Nernst coefficient $N_s$ versus temperature at different $\alpha$ in Fig. \ref{Fig5}.
$N_s$ exhibits the oscillation at low temperature (like Fig. \ref{Fig4}(a)), while it becomes smooth and approximately proportional to the temperature $\mathcal{T}$ at relatively high temperature.
As mentioned in Sec. \ref{sec3}, the spin splitting Nernst coefficient comes from the derivative of $\Delta T_{23}-\Delta T_{21}$ near $E_F$,
and the effective range of the integral in Eq. (\ref{E8}) is proportional to the temperature $\mathcal{T}$. 
As the temperature increases from $0$, the effective integral range expands.
As this range covers one peak of $\frac{d(\Delta T_{23}-\Delta T_{21})}{dE}$ (about $\mathcal{T}=0.004$), the spin splitting Nernst coefficient exhibits a peak as shown in Fig. \ref{Fig5}.
With the further increase of temperature, more and more electrons participate in the thermoelectric transport. 
As a result,
the high temperature suppresses the oscillation and enhances the magnitude of the spin splitting Nernst effect \cite{xing_Nernst_2009}.
Besides, Fig. \ref{Fig5} also shows that in certain parameter range, a system with a larger altermagnet strength $\alpha$ has a more pronounced spin splitting Nernst effect.

\begin{figure}[!htb]
\centerline{\includegraphics[width=\columnwidth]{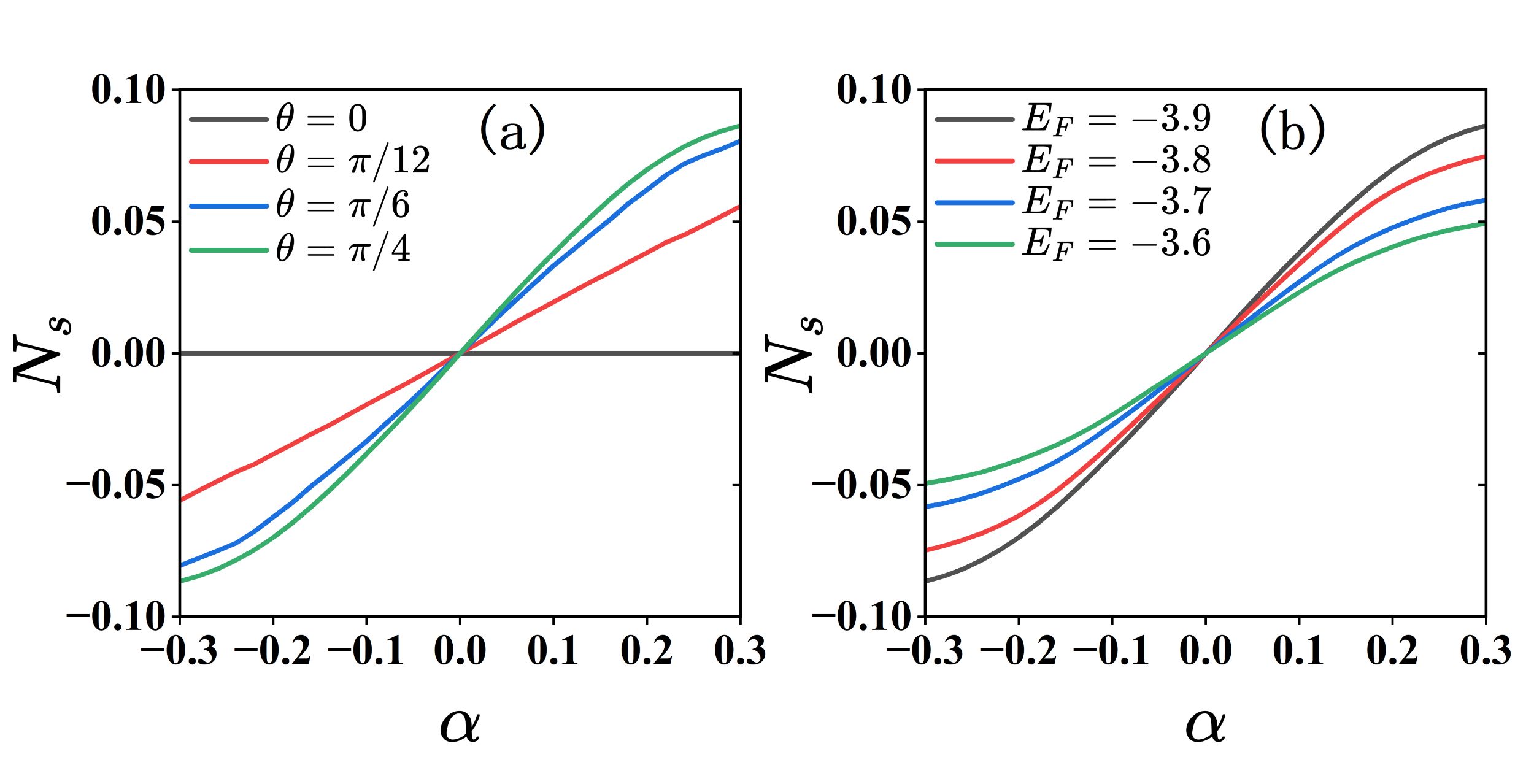}}
\caption{$N_s$ versus $\alpha$ for different $\theta$ and different $E_F$ in (a) and (b), respectively. The parameters $\mathcal{T}=0.06$ and $L=60$ for (a) and (b), $E_F$=-3.9 for (a), and $\theta=\pi/4$ for (b). }\label{Fig6}
\end{figure}

Next, we study the effect of the strength of altermagnet $\alpha$ on the spin splitting Nernst coefficient $N_s$. Fig. \ref{Fig6}(a) shows the spin splitting Nernst coefficient $N_s$ versus $\alpha$ at different $\theta$ and Fig. \ref{Fig6}(b) shows $N_s$ versus $\alpha$ at different $E_F$.
In both figures, the spin splitting Nernst coefficient $N_s$ is almost proportional to $\alpha$. 
When $\alpha$ is zero, the Fermi surface is round.
As discussed in Sec. \ref{sec2}, in average, there is not a transverse velocity to drive the electrons transversely, leading to the absence of the spin splitting Nernst effect.
When $\alpha$ is nonzero, the Fermi surface is deformed to ellipse, and the nonzero transverse velocity drives the electrons with opposite spins to opposite directions transversely.
With the increase of $\alpha$, the anisotropy of elliptical Fermi surface is lifted \cite{sun_Andreev_2023}, and the electrons are transversely driven more intensely.
The result is also consistent with our discussion in Sec. \ref{sec2}, where the spin splitting Nernst coefficient is proportional to $\alpha \sin2\theta$.
In addition, $N_s$ is an odd function of $\alpha$ with $N_s (\alpha)=-N_s (-\alpha)$, which can be verified by the symmetry analysis.
For the spin exchange operator $\sigma_x$, the Hamiltonian satisfies $\sigma_x H(\alpha,\theta) \sigma_x^{-1}=H(-\alpha,\theta)$, the operation only exchanges spin, so the transverse spin current is inversed.
Thus we obtain
\begin{align}
N_s(\alpha,\theta) = -N_s(-\alpha,\theta).
 \label{newE10}
\end{align}
Fig. \ref{Fig6}(a) shows the strong dependence of $N_s$ on the angle $\theta$.
At $\theta=0$, $N_s$ is zero.
With the increase of $\theta$ from zero to $\pi/4$, the spin splitting Nernst effect is more pronounced.
On the other hand, Fig. \ref{Fig6}(b) shows that the increase of Fermi energy $E_F$ decreases the spin splitting Nernst effect, consistent with the result in Fig. \ref{Fig4}.

\begin{figure}[!htb]
\centerline{\includegraphics[width=\columnwidth]{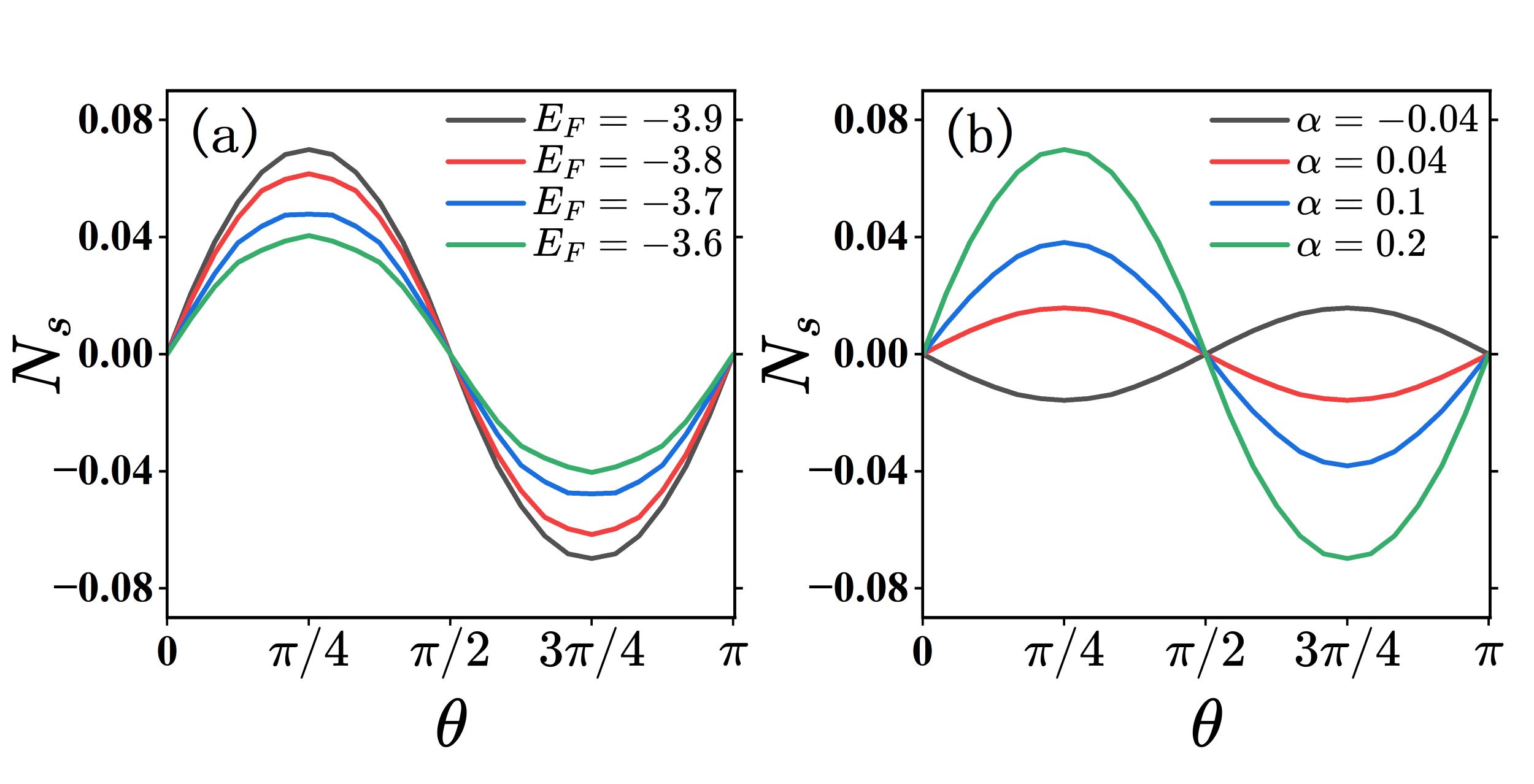}}
\caption{$N_s$ versus $\theta$ for different $E_F$ and different $\alpha$ in (a) and (b), respectively. The parameters $\mathcal{T}=0.06$ and $L=60$ for (a) and (b), $E_F=-3.9$ for (b) and $\alpha=0.2$ for (a). }\label{Fig7}
\end{figure}

Now we study how the angle $\theta$ affects the spin splitting Nernst effect.
We plot the spin splitting Nernst coefficient $N_s$ versus $\theta$ at different $E_F$ in Fig. \ref{Fig7}(a) and at different $\alpha$ in Fig. \ref{Fig7}(b).
The curves are all consistent with the $\sin 2\theta$ relation obtained in Sec. \ref{sec2}.
When the $\theta$ is the integer multiple of $\pi/2$, the $N_s$ is zero, corresponding to that the major axis of the elliptical Fermi surfaces coincide with $x$ or $y$ axis, and the Fermi surface is symmetric about $x$ axis.
In this condition, for both spin-up and spin-down, the electron currents with opposite transverse motions cancel out.
On the other hand, when the $\theta$ is not the integer multiple of $\pi/2$, the elliptical Fermi surface is not symmetric about the $x$ axis, and the spin splitting Nernst effect is induced under a longitudinal thermal gradient.
When $\theta=\pi/4$, the spin splitting Nernst effect is the most pronounced, meaning the thermal diffusing electrons are driven by the largest transverse velocity when the angle between crystalline axis and longitudinal direction is $\pi/4$.
In addition, from the curves in Fig. \ref{Fig7}, the angle period is $\pi$, also with the antisymmetry center  $\pi/2$ and the symmetry axis $\pi/4$.
To explain them, we derive the symmetry relation from the spin-exchange operation $\sigma_x$ and the rotation operation about the $x$ axis $R_x (\pi)$ \cite{cheng_Spintriplet_2021, fang_Bulk_2012, yan_Anomalous_2020}.
The operation $R_x (\pi)$ only reverses the spatial transverse direction, and has no influence on spin.
We first just list the effects of the operations on the Hamiltonian
\begin{align}
    &\sigma_x H(\alpha,\theta) \sigma_x^{-1}=H(-\alpha,\theta), \nonumber \\
    &R_x (\pi)H(\alpha,\theta) R_x (\pi)^{-1}=H(-\alpha,\frac{\pi}{2}-\theta)    .     \label{E10}
\end{align}
From Eq. (\ref{E5}), the Hamiltonian itself satisfies
\begin{align}
    H(\alpha,\theta)=H(-\alpha,\theta+\frac{\pi}{2})
    =H(\alpha,\theta+\pi).    \label{E11}
\end{align}
The spin splitting Nernst coefficients of the two equivalent systems are equal to each other, therefore, $N_s(\alpha,\theta)=N_s(-\alpha,\theta+\pi/2)
=N_s(\alpha,\theta+\pi)$,
indeed with a $\pi$ period.
In addition, because $N_s$ is an odd function of $\alpha$ from Eq.(\ref{newE10}), $N_s(\alpha,\theta)=N_s(-\alpha,\theta+\pi/2)=-N_s(\alpha,\theta+\pi/2)$.
Furthermore, considering the joint operation $\mathcal{Y}= \sigma_x R_x (\pi)$, under the operation $\mathcal{Y}$ and by using Eq.(\ref{E10}), the Hamiltonian satisfies
\begin{align}
    \mathcal{Y}H(\alpha,\theta)\mathcal{Y}^{-1}=H(\alpha,\frac{\pi}{2}-\theta)    .    \label{E12}
\end{align}
First, the rotation operation $R_x(\pi)$ inverses the transverse direction. $N_s$ is turned to $-N_s$.
Moreover, we have mentioned previously that spin-exchange operation $\sigma_x$ turns $N_s$ to $-N_s$.
Under the joint operation, the sign of $N_s$ keeps unchanged.
Combining with the change of the Hamiltonian under the joint operation in Eq. (\ref{E12}) that $\theta$ turns to $\pi/2-\theta$,
$N_s$ has the relation $N_s (\alpha,\theta)=N_s (\alpha,\pi/2-\theta)$,
which shows the symmetry axis $\pi/4$. Along with the relation $N_s (\alpha,\theta)=-N_s (\alpha,\theta+{\pi/2})$, we obtain $N_s (\alpha,\pi/2-\theta)=-N_s (\alpha,{\pi/2} +\theta)$
, which shows the symmetry center $\pi/2$.
This angle relations are universally consistent with the results in Figs. \ref{Fig7}(a, b).
In Fig. \ref{Fig7}, within certain range, $N_s$ is larger with the decrease of $E_F$ in the whole $\theta$ space.
In addition, a system with a larger altermagnet strength $\alpha$ has a larger $N_s$, and the relation $N_s(\alpha)=-N_s(-\alpha)$ is still tenable at different $\theta$.

\begin{figure}[!htb]
	\centerline{\includegraphics[width=\columnwidth]{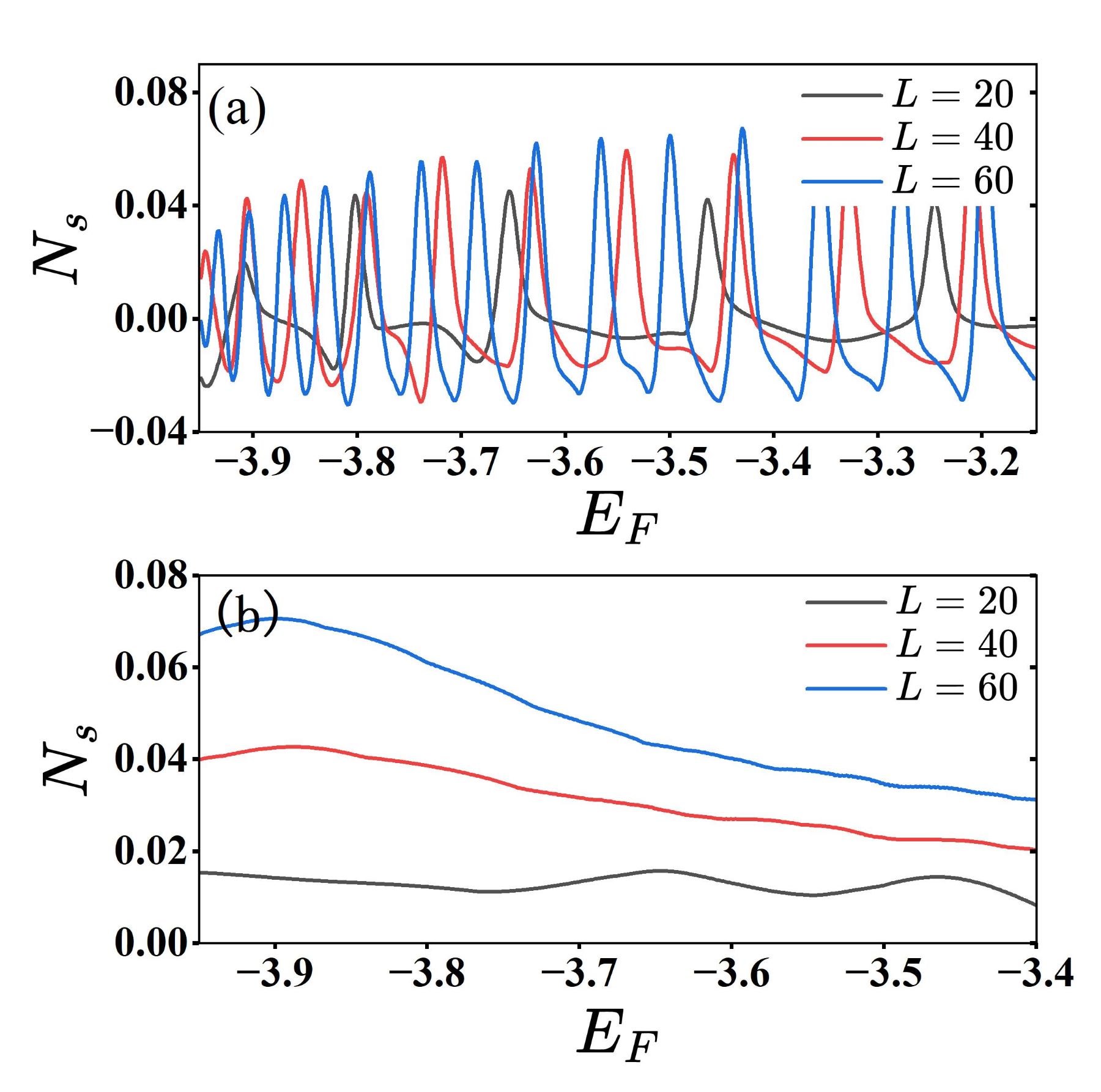}}
	\caption{$N_s$ versus $E_F$ for different sizes. In panel (a) $\mathcal{T}=0.004$ and in panel (b) $\mathcal{T}=0.06$. The other parameters are $\alpha=0.2$ and $\theta=\pi/4$ in (a) and (b).}\label{Fig8}
\end{figure}

We also study the effect of the size of the sample $L$.
Figs. \ref{Fig8}(a, b) show the spin splitting Nernst coefficient $N_s$ versus $E_F$ with different sizes of the sample at $\mathcal{T}=0.004$ and $\mathcal{T}=0.06$, respectively.
At a low temperature in Fig. \ref{Fig8}(a), a series of oscillatory peaks appear. With the larger size $L$, adjacent peaks become closer.
That is because the spacing of adjacent subbands is diminished \cite{datta_Electronic_1995}, the sudden jumps of transmission coefficient become dense, leading to the decrease of the spacing of adjacent peaks.
In Fig. \ref{Fig8}(b), at room temperature with the oscillation damped, the spin splitting Nernst coefficient $N_s$ curve becomes smooth.
Although there is still a little oscillation for a small size $L=20$, the increase of $L$ to $60$ further smooths the curve due to the dense subbands.
This also means the increase of the size weakens the quantum effect of our calculations.
On the other hand, $N_s$ is enhanced with a larger size of the sample, and $N_s$ for $L=60$ is nearly three times as large as that for $L=20$.
That is because at the same energy, there are more transport modes for thermally driven electrons to move transversely with the increase of the sample size.

\section{\label{sec5} Discussion and Conclusion}

Above we study the effect that an $x$-direction temperature difference drives a $y$-direction spin current, and the spin splitting Nernst coefficient in Eq. (\ref{E8}) can be labelled as $N_{s,xy}$.
For the $y$-direction temperature difference driving an $x$-direction spin current, it corresponds to the label $N_{s,yx}$.
We propose that in altermagnet, the $xy$-response and $yx$-response spin splitting Nernst coefficients are equal with $N_{s,xy}=N_{s,yx}$.
We can prove this by deriving how an $x$-direction spin current is induced by a $y$-direction temperature gradient.
By setting the temperatures of the leads as $\mathcal{T}_4=\mathcal{T}+\frac{\Delta \mathcal{T}}{2}$, $\mathcal{T}_2=\mathcal{T}-\frac{\Delta \mathcal{T}}{2}$, $\mathcal{T}_1=\mathcal{T}_3=\mathcal{T}$, 
we can derive the $N_{s,yx}=J_{3s}/\Delta \mathcal{T}$, analogous to Eq. (\ref{E8})
\begin{align}
    N_{s,yx}=\frac{1}{4\pi} \int dEf_0 (1-f_0) \left [\frac{(E-E_f )}{k_B \mathcal{T}^2} (\Delta T_{32}-\Delta T_{34} )\right ].    \label{E13}
\end{align}
Here, $\Delta T_{32}=T_{3\uparrow,2}-T_{3\downarrow,2},\Delta T_{34}=T_{3\uparrow,4}-T_{3\downarrow,4}$.

Consider the symmetry relations by the spin exchange operation $\sigma_x$ and rotation operation by $\pi/2$ about the $z$ axis $R_z(\pi/2)$ , one gets
\begin{align}
    & \sigma_x H(\alpha,\theta)  \sigma_x^{-1}=H(-\alpha,\theta), \nonumber \\
    &R_z (\frac{\pi}{2})H(\alpha,\theta) R_z (\frac{\pi}{2})^{-1}=H(-\alpha,\theta)    .         \label{E14}
\end{align}
Under the rotation operation,
$T_{3\sigma,2}$ is changed to $T_{2\sigma,1}$, $T_{3\sigma,4}$ is changed to $T_{2\sigma,3}$.
$\Delta T_{32}-\Delta T_{34}$ is changed to $\Delta T_{21}-\Delta T_{23}$.
Under the spin exchange operation,
$T_{p\sigma,q}$ is changed to $T_{p\overline{\sigma},q}$,
$\Delta T_{pq}$ is changed to $-\Delta T_{pq}$ ($p, q$ are the indices of leads).
Under the joint operation, $\Delta T_{32}-\Delta T_{34}$ is changed to $-\Delta T_{21}+\Delta T_{23}$, meanwhile, the Hamiltonian keeps unchanged,
$\sigma_xR_z (\frac{\pi}{2})H(\alpha)R_z (\frac{\pi}{2})^{-1}\sigma_x^{-1}=H(\alpha)$.
Thus, $\Delta T_{32}-\Delta T_{34}=\Delta T_{23}-\Delta T_{21}$.
According to Eqs. (\ref{E8},\ref{E13}), $N_{s,xy}=N_{s,yx}$ is obtained.

In Ref. \cite{gonzalez-hernandez_Efficient_2021},
it was proposed that for a time-reversal-even source, such as spin-orbit coupling, the corresponding spin transverse conductivity satisfies $\sigma_{xy}=-\sigma_{yx}$.
But for a time-reversal-odd source, such as altermagnetism studied by us, the corresponding spin transverse conductivity satisfies $\sigma_{xy}=\sigma_{yx}$.
Here in our study, similarly, the spin splitting Nernst effect is induced by the time-reversal-odd source, altermagnetism.
Therefore, the coefficient also satisfies $N_{s,xy}=N_{s,yx}$, which is different from the conventional spin Nernst effect where they are opposite with $N_{s,xy}=-N_{s,yx}$.

In conclusion, we propose the spin splitting Nernst effect in altermagnet from electron transverse velocity.
We give the physical mechanism that spin-up and spin-down electrons oppositely split in transverse direction under the longitudinal thermal gradient.
The spin splitting Nernst effect is verified by calculations in a four-terminal device.
With the aid of the tight-binding Hamiltonian and the nonequilibrium Green's function method, we obtain the nonzero spin splitting Nernst coefficient.
We also systematically study the parametric dependency of spin splitting Nernst effect, and perform corresponding symmetry analysis.
It is worth noting that such spin splitting Nernst effect does not require spin-orbit coupling and net macroscopic magnetism.
The results above are beneficial to the further development of spintronics
and the design of thermoelectric devices.

\begin{acknowledgments}
We thank Yu-Chen Zhuang for fruitful discussions.
This work was financially supported
by the National Key R and D Program of China (Grant Nos. 2024YFA1409002),
NSF-China (Grants No. 11921005 and No. 12374034),
and the Innovation Program for Quantum Science and
Technology (No. 2021ZD0302403). The computational
resources are supported by the High-Performance
Computing Platform of Peking University.
\end{acknowledgments}

\bibliography{refer.bib}
\end{document}